\documentclass[
 reprint,
 superscriptaddress,
 apl,
 amsmath,amssymb,
 prl
 ]{revtex4-2}
\usepackage[utf8]{inputenc}
\usepackage{graphicx}
\usepackage{dcolumn}
\usepackage{bm}
\usepackage{xcolor}
\usepackage{amsmath}
\usepackage[T1]{fontenc}
\DeclareUnicodeCharacter{2212}{-}

\begin{document}

\title{Optimization of submicron Ni/Au/Ge contacts to an AlGaAs/GaAs two-dimensional electron gas}

\author{Matthew Mann}

 \affiliation{Elmore Family School of Electrical and Computer Engineering, Purdue University, West Lafayette, Indiana 47907, USA}%

\author{James Nakamura}

 \affiliation{Department of Physics and Astronomy, Purdue University, West Lafayette, Indiana 47907, USA}

\author{Shuang Liang}
 \affiliation{Department of Physics and Astronomy, Purdue University, West Lafayette, Indiana 47907, USA}

\author{Tanmay Maiti}

 \affiliation{Department of Physics and Astronomy, Purdue University, West Lafayette, Indiana 47907, USA}
\author{Rosa Diaz}

 \affiliation{Birck Nanotechnology Center, Purdue University, West Lafayette, Indiana 47907, USA}
 
\author{Michael J. Manfra}

\affiliation{Elmore Family School of Electrical and Computer Engineering, Purdue University, West Lafayette, Indiana 47907, USA}
\affiliation{Department of Physics and Astronomy, Purdue University, West Lafayette, Indiana 47907, USA}
\affiliation{Birck Nanotechnology Center, Purdue University, West Lafayette, Indiana 47907, USA} 
\affiliation{School of Materials Engineering, Purdue University, West Lafayette, Indiana 47907, USA}
\affiliation{Microsoft Quantum Lab West Lafayette, West Lafayette, Indiana 47907, USA}

\begin{abstract}
We report on fabrication and performance of submicron Ni/Au/Ge contacts to a two-dimensional electron gas in an AlGaAs/GaAs heterostructure. Utilizing scanning transmission electron microscopy, energy dispersive x-ray spectroscopy, and low temperature electrical measurements we investigate the relationship between contact performance and the mechanical and chemical properties of the annealed metal stack. Contact geometry and crystallographic orientation significantly impact performance. Our results indicate that the spatial distribution of germanium in the annealed contact plays a central role in the creation of high transmission contacts. We characterize the transmission of our contacts at high magnetic fields in the quantum Hall regime. Our work establishes that contacts with area 0.5\(\mu\)m\(^2\) and resistance less than 400\(\Omega\) can be fabricated with high yield.
\end{abstract}
\maketitle

\section{\label{sec:level1}}
The two-dimensional electron gas (2DEG) formed in modulation-doped AlGaAs/GaAs heterostructures is a workhorse for experimental exploration of mesoscopic physics, the fractional quantum Hall effect, and construction of solid-state qubits \citep{Sohn, Beenakker, Weis, Willet, Nakamura, Hartman}. Establishing electrical contact to the buried 2DEG is hindered by the Schottky barrier that forms at the metal-semiconductor interface \cite{BaslauSchottky, Waldrop, KoopEtAl} and the remoteness of the 2DEG from the surface. A common approach to make ohmic contacts is thermal annealing a metal stack of Ni/Au/Ge that has been deposited on the surface \citep{Braslau, Kurochka, Heiblum, Rai, Kane, Ahlswede}. By decreasing the contact size to submicron scale, galvanic connections can be placed near the active region of mesoscopic devices facilitating measurement of a wide range of interesting physics including anyon interference and the multichannel Kondo effect \citep{Kundu, PierreKondo, PouseQuantumSim, GoktasThesis}. A few published studies have reported annealing parameters and ohmic metal stack variations pursuant to generation of micron-scale contacts \citep{KoopEtAl, Goktas, Christou.1979, Buhlmann, Procop, Kim, Brown}. Despite these efforts, creation of high-yield, low resistance micron-scale contacts that remain highly transmitting in the quantum Hall regime remains an outstanding challenge. Our study builds upon previous works by explicating a high-yield fabrication process, and provides analysis of the impact of contact geometry, crystallographic orientation, and spatial distribution of metallic elements in the post-annealed stack that determine the functionality of small ohmic contacts. 

To investigate the quality of our micron-scale ohmic contacts we fabricate multiple devices using two distinct geometries: traditional high aspect ratio Hall bars and a second mesa design that minimizes the contribution of the 2DEG to our resistance measurements by placing one large area ohmic contact near seven micron-scale contacts with simultaneous reduction of the area of 2DEG between contacts. This second device geometry is shown schematically in the inset in Fig. 2(a). Seven micron-scale contacts are aligned to the [01\(\overline{1}\)]-oriented mesa edge spaced 20\(\mu\)m away from the single large ohmic drain contact. For the Hall bars, 2mm long $\times$ 150$\mu$m wide rectangular mesas are defined that are oriented parallel to the [011] or the [01\(\overline{1}\)] crystallographic direction of the GaAs substrate. We place six micron-scale ohmic contacts along each edge of the Hall bar with large ohmics at each end of the mesa serving as source and drain contacts; the orientation, size, and geometry of the micron-scale contacts are varied in our experiments. As shown in Fig. 1(c,d), the contacts are either circular or rectangular. Our devices are fabricated on a modulation-doped AlGaAs/GaAs heterostructure grown by molecular beam epitaxy with a 2DEG positioned 91nm below the surface. As shown in Fig. 1(a), the active region consists of a 500nm GaAs layer, a 45nm undoped AlGaAs spacer layer, a 11.3nm silicon-doped AlGaAs region, a 28nm undoped AlGaAs layer, and a 7nm GaAs capping layer. The wafer presented in this study has a 2DEG electron density of 1.9$\times$10$^{11}$ cm$^{-2}$ and mobility of 3.2$\times$10\(^6\) cm\(^2\)/Vs at T=0.3K.

\begin{figure}
\centering 
\includegraphics[width=0.5\textwidth,angle=0]{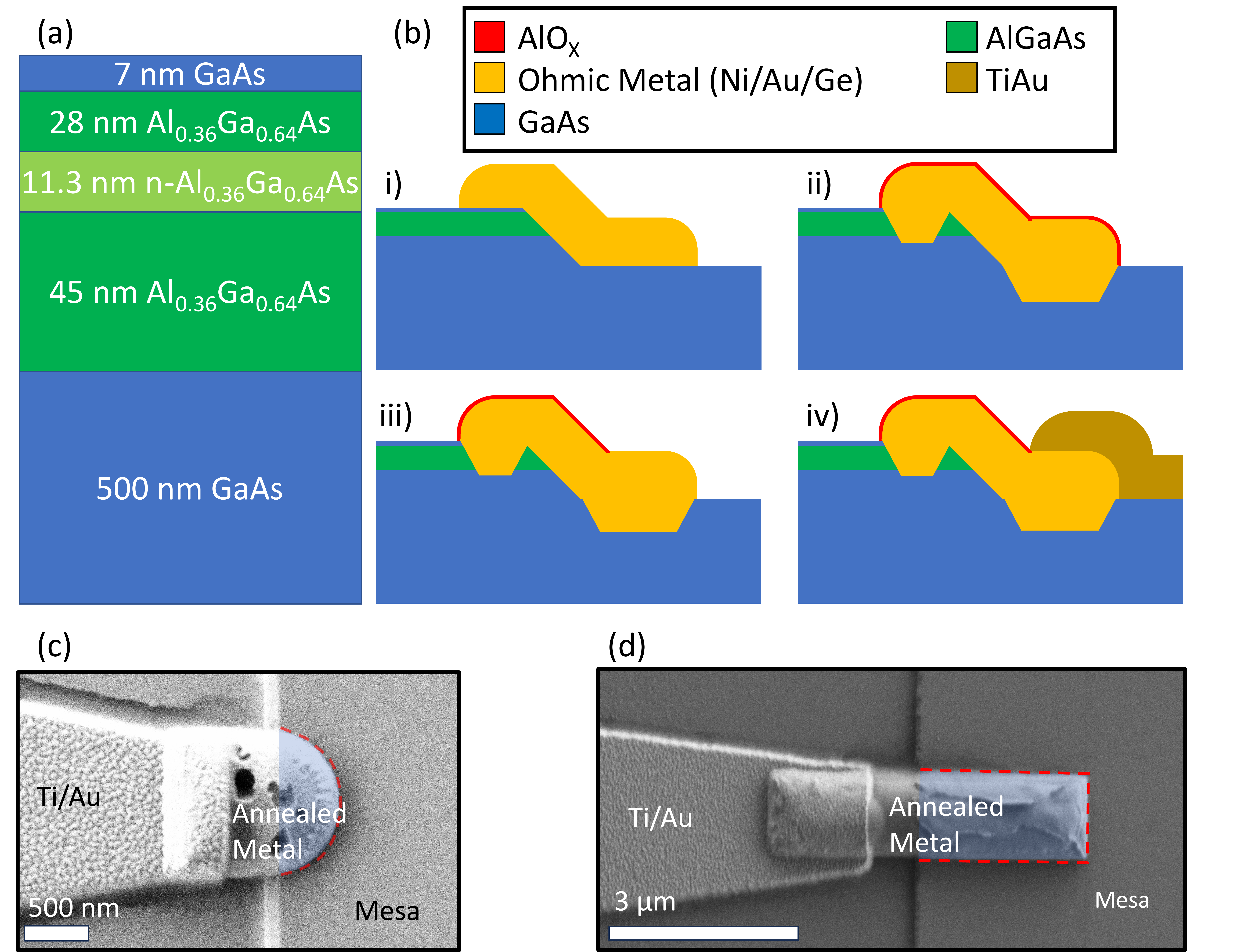}	
\caption{(a) Schematic of the AlGaAs/GaAs heterostructure used in these experiments. All devices were fabricated from a single wafer.
(b) Schematic of fabrication flow. i) Deposited metal stack (yellow). ii) Metal is annealed. Aluminum in the AlGaAs barrier (green) migrates from heterostructure to metal surface where it forms a surface oxide (red). iii) Aluminum oxide is etched away with TMAH solution. iv) TiAu (brown) deposited.
(c) SEM image of circular micron-scale ohmic contact. False colors indicate area (blue) and perimeter (red).
(d) SEM image of rectangular micron-scale ohmic contact.} 
\end{figure}

Our experiments have identified several critical processing steps for the creation of robust and low resistance contacts. The crucial steps include: cleaning of the semiconductor surface after lithographic definition of the location for micron-scale contacts, but prior to deposition of the ohmic metal stack; organic residue removal immediately after contact deposition but prior to the high temperature anneal; and oxide removal before deposition of the metal that forms the fan-out from the annealed micron-scale ohmic to the bond pads. A detailed description of our micron-scale ohmic contact fabrication flow is given in the supplementary material.

The results of our first controlled variation study are displayed in Fig. 2(a). In this experiment, the cleaning procedure immediately preceding the ohmic metal deposition was investigated. All contacts were annealed at 435\(^\circ\)C for 2 minutes. We alternatively used an oxygen plasma only treatment, an oxygen plasma treatment followed by an additional HCl etch prior to loading into the metal evaporation system, and for a third sample we eliminated all cleaning treatments after pattern definition. For these experiments, we used the device geometry shown in inset to Fig. 2(a) with circular contacts aligned along the [01\(\overline{1}\)]-oriented mesa edge. Use of a 15s oxygen plasma treatment resulted in the average resistance of 356\(\pm\)24\color{black}\(\Omega\) for 0.5\(\mu\)m\(^2\) contacts. The error listed is the standard error of the mean calculated by dividing the standard deviation of the resistance by the square root of the number of contacts measured. No cleaning treatment immediately prior to ohmic metal deposition resulted in an average resistance of 411\(\pm\)36\color{black}\(\Omega\) for 0.5\(\mu\)m\(^2\) contacts. Interestingly, the average resistance was 800\(\pm\)95\color{black}\(\Omega\) when we included a 20s HCl etch after the oxygen plasma treatment immediately prior to loading the sample into the metal deposition chamber. Cumulatively these results suggest that cleaning procedures designed to remove organic residue and surface oxides that may etch or damage the top layers of semiconductor can negatively impact the resistance of micron scale contacts. Nevertheless, the yield of working contacts remained 100\(\%\) regardless of the cleaning procedure variations used to generate these data sets. The data of Fig. 2(a) establishes that submicron area ohmic contacts with average resistance less than 400\(\Omega\) can be fabricated with high yield.

For the data displayed in Fig. 2(b) we used both circular and rectangular contact geometries and an on-mesa contact area of 1$\mu$m$^2$. We varied only the annealing time and temperature while keeping other parameters of the fabrication process fixed. The data in Fig. 2(b) indicates that contact geometry also plays an important role in contact yield and variations. Fig. 2(b) shows when the annealing parameters are varied, the resistance of the rectangular contacts fluctuates significantly, while the circular contacts maintain a consistently low value of average resistance with less run-to-run variation. For contacts with circular geometry, variations in anneal temperature of 10\(^\circ\)C and anneal time of one minute do not dramatically influence the average contact resistance. 
\begin{figure}
	\centering 
	\includegraphics[width=0.5\textwidth, angle=0]{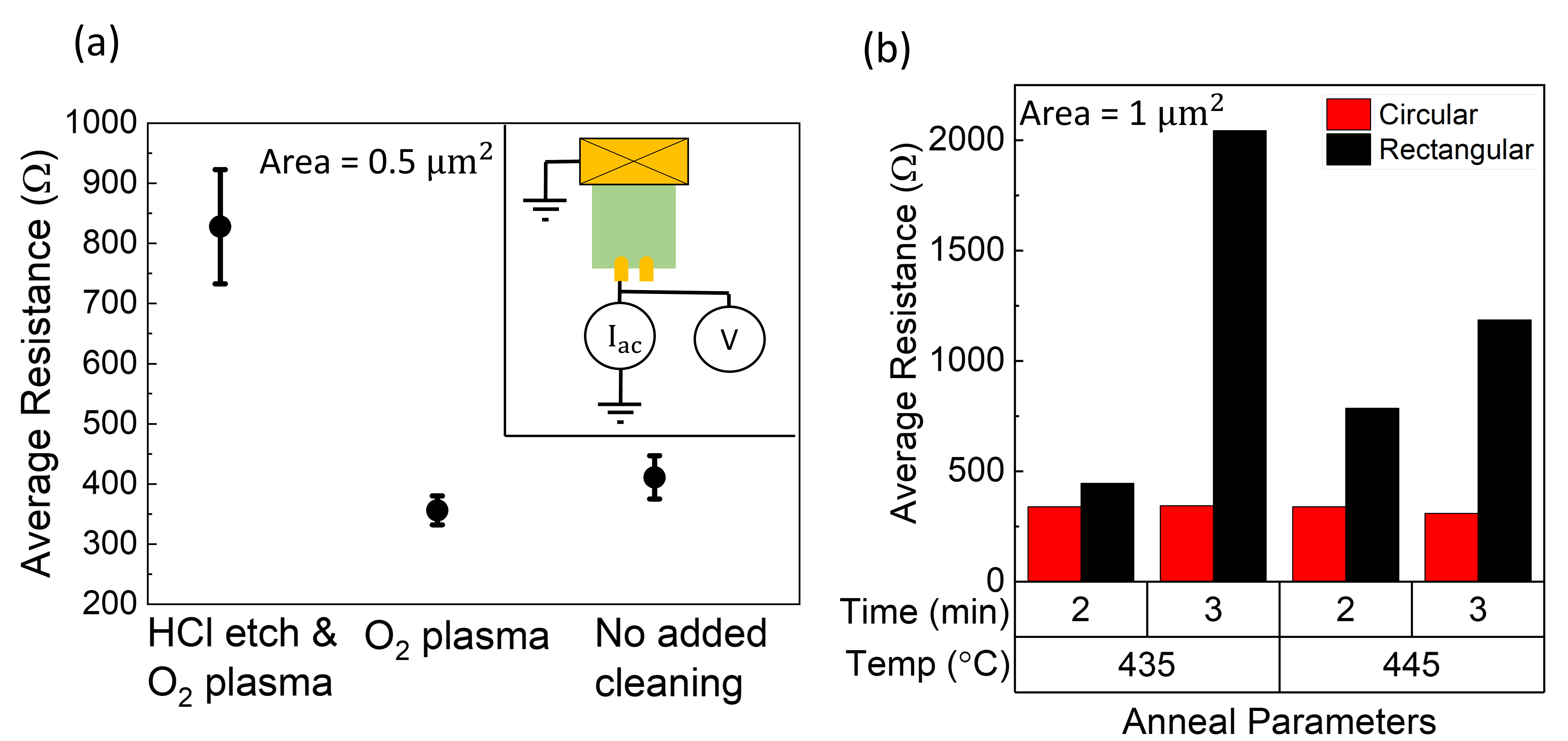}	
	\caption{(a) Effect of different cleaning procedures prior to ohmic metal stack deposition. Inset shows device and measurement configuration with reduced 2DEG contribution. These measurements were conducted at T=4.2K.
(b) Impact of variation of annealing parameters on contacts with rectangular and circular geometries. All contacts are oriented along the [01\(\overline{1}\)] direction and have an area of 1\(\mu\)m\(^2\). The contacts were fabricated simultaneously to minimize variations other than anneal conditions. Resistance is measured at T=4.2K.} 
\end{figure}

For the data reported in Fig. 3(b), the standard Hall bar geometry shown schematically in Fig. 3(a) is used. For this fabrication run, the semiconductor surface where we deposit the ohmic metal stack is cleaned with an oxygen plasma and an additional 30 second {\it in-situ} argon ion mill in the metal deposition chamber immediately prior the deposition of the ohmic metal stack. We note that this argon ion milling step {\it did not improve} the quality of our contacts. In fact the contact resistance was higher than the best results achieved at each contact size.

Our initial electrical characterization is carried out at T=4.2K. For the Hall bars, we inject a 1 \(\mu\)A low-frequency (\(<\) 100 Hz) current into a micron-scale contact and ground all other contacts as shown in Fig. 3(a). We measure the voltage at the micron-scale contact using standard low-frequency lock-in techniques and convert the result to a resistance. We account for the small, but finite, resistance of the 2DEG in the Hall bar and line resistances in the cryostat. We vary the contact size and contact orientation relative to the crystallographic axes of the GaAs substrate. In Fig. 3(b), we take the measured resistance values for each circular contact and calculate the average for each unique parameter combination. The standard error of the mean is represented by the bars associated with each data point in Fig. 3(b). For a contact area of 0.5\(\mu\)m\(^2\), this fabrication run yields an average resistance of approximately 800\(\pm\)61\color{black}\(\Omega\) for circular contacts along the mesa edge parallel to the [01\(\overline{1}\)] direction. The yield of working 0.5\(\mu\)m\(^2\) contacts was 100\(\%\). The contact resistance does depend on alignment to a specific crystallographic axis; this dependence may be associated with the contact's exposure to crystallographic planes of GaAs other than (001) on the sidewalls of the mesa edges. For both orientations, the wet etch of the GaAs mesa creates sloped sidewalls; the deposited metal stack remains conformal. We also note that for both orientations the average resistance scales inversely with contact area. In Fig. 3(b) functions corresponding to 1/area are displayed and appear to account for the trends in the data.
\begin{figure}
\includegraphics[width=0.5\textwidth, angle=0]{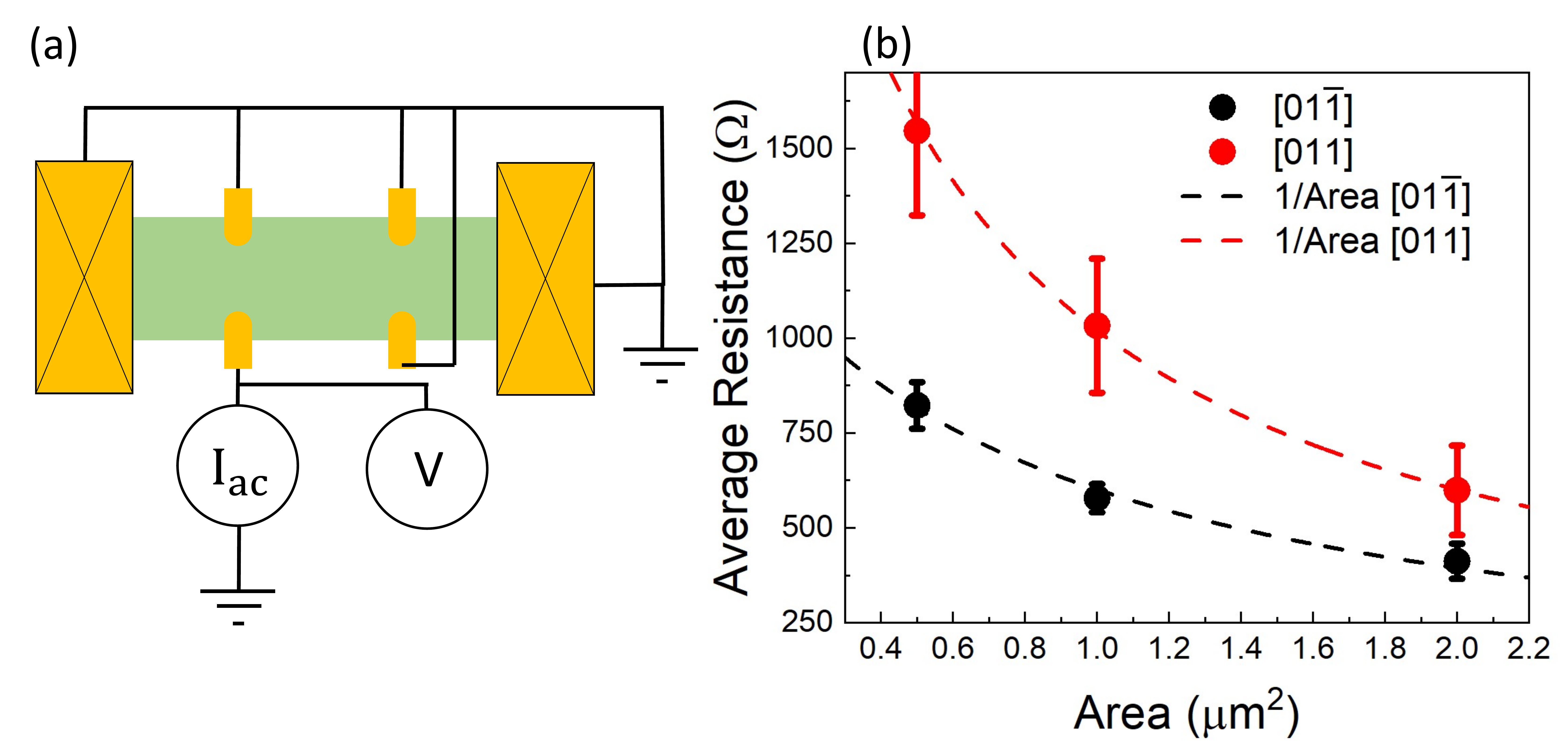}	
\caption{
(a) Measurement configuration to assess contact resistance in the Hall bars. Current is injected into a single contact, while all others are grounded.
(b) Effect of contact area and crystallographic orientation for circular contacts on a Hall bar. Contact resistance is measured at T=4.2K} 
\end{figure}

While our yield remains 100\(\%\), we still observe a rise in resistance as contact area decreases, as shown in Fig. 3(b). In the literature, this increase has been attributed to reduced perimeter connecting the metal to the surrounding 2DEG \citep{KoopEtAl, Taylor, Goktas}. We also observe that the orientation of the contact relative to the crystallographic axis of the GaAs substrate affects the resistance. As shown in Fig. 1(a) our contacts are aligned to and overlap the edge of the mesa. This edge will either be parallel to the [011] or the [01\(\overline{1}\)] direction. When aligned to the [01\(\overline{1}\)] direction, the average contact resistance and the variation decreases. The average resistance drops by as much as 40\(\%\) and the standard deviation is halved when a contact is aligned along the [01\(\overline{1}\)] direction while all other parameters are kept constant. 

The observed orientation dependence has been reported in previous works \citep{Kamada, Goktas, Kane}. The origin of the orientation dependence has not been conclusively established, but an explanation has been put forth in literature that focuses on the anisotropy of metal diffusion and the interaction between the grains of metallic compounds formed during the anneal and the semiconductor crystal lattice \citep{Kurochka, Christou.1979, Kamada, Kuan, Rai}. During the anneal, nickel diffuses into the substrate and interacts with the germanium and the arsenic to form NiAs and NiGeAs grains. Germanium migration into the semiconductor is further enhanced as germanium transfers from the NiGe grains to the NiGeAs grains \citep{Kuan, KoopEtAl}. Once germanium penetrates far enough into the semiconductor, it will heavily dope the area around the metal contact and create a highly conductive connection to the 2DEG. In addition, it has been observed that the crystal lattice of the NiAs and NiGeAs grains will orient themselves along specific crystallographic directions of the host lattice \citep{Christou.1979, Kuan}. Since germanium diffusion is assisted by the presence of nickel, it is then also influenced by the orientation of the NiGeAs grains. Germanium is ultimately responsible for the heavy n-type doping of the surrounding semiconductor facilitating good contact to the 2DEG, thus explaining the orientation dependence of the contact resistance.

Our data indicates that contact geometry also plays an important role in contact yield and variations. Fig. 2(b) shows that our circular contacts have extremely consistent resistance values compared to contacts with rectangular shape. Evidently, circular contacts allow the metal to diffuse more uniformly and to sample favorable crystallographic axis alignment, ensuring a more optimal distribution of germanium in the heavily doped semiconductor region adjacent to the 2DEG. The circular contact still shows an orientation dependence, as demonstrated by Fig. 3(a). This dependence is associated with the contact's exposure to the mesa edge. Thus, the optimal contact is created with a circular geometry and aligned to the [01\(\overline{1}\)] crystallographic direction.\newline

We further characterize the electrical properties of our contacts by measuring transmission in the quantum Hall regime at filling factors $\nu=1$ and $\nu=2$ at T=0.3K. A 10\(\mu\)V AC voltage is applied to the large source contact. Current is drained at the micron-scale contact under test and at the large drain contact at the end of the Hall bar. In the context of Landauer-B\"uttiker formalism, transmission is defined as the probability of an electron flowing from one terminal to another terminal point. The current can be expressed as the current carried by the mode times the average transmission probability of the contact under test. 
\begin{eqnarray}
I_T = (e^2/h)\times M \times T \times (V_i - V_j)
\end{eqnarray}
\begin{eqnarray}
I_R = (e^2/h) \times M \times (1 - T) \times (V_i - V_j)
\end{eqnarray}
\begin{eqnarray}
T = I_T / (I_T + I_R)
\end{eqnarray}
Here $I_T$ refers to the current drained through the micron scale ohmic contact and $I_R$ is the current that is drained through the large drain contact at the end of the Hall bar as shown in Fig. 4(a). $M$ corresponds to the number of edge modes, $T$ is the transmission of the micron-scale contact, and \(V_i\) and \(V_j\) are the voltages at the contacts under consideration. Note that since current is carried by edge modes in the quantum Hall regime, this analysis assumes all edge modes have the same transmission for a given contact. If the micron-scale contact is perfectly transmitting, the entirety of the impinging edge mode current will drain through it. Any current that drains through the large drain contact at the end of the Hall bar is referred to as reflected current. The transmission can be expressed as ratio between the current flowing out through the micron-scale contact and the sum of the reflected and transmitted current, i.e. the total current. As shown in Fig. 4(c) our contacts have high transmission at the \(\nu\)=1 and \(\nu\)=2 quantum Hall states. Transmission at $\nu=1$ is above 99$\%$ for the 1\(\mu\)m\(^2\) contacts. As the net transmission is slightly less at $\nu=2$, we may speculate that the transmission of the slower, inner edge mode at $\nu=2$ is reduced in comparison with the outermost mode. Once we have measured the transmission of all the micron-scale ohmics, we change the measurement configuration to source and sink the current through the large-area contacts and use the micron-scale contacts as voltage probes for magnetotransport measurements. Fig. 4(d) shows the Hall resistance and longitudinal resistance measured using 0.5\(\mu\)m\(^2\) contacts and measured at T=0.3K in magnetic field up to B=8.5T. 
\begin{figure}
\centering 
\includegraphics[width=0.5\textwidth, angle=0]{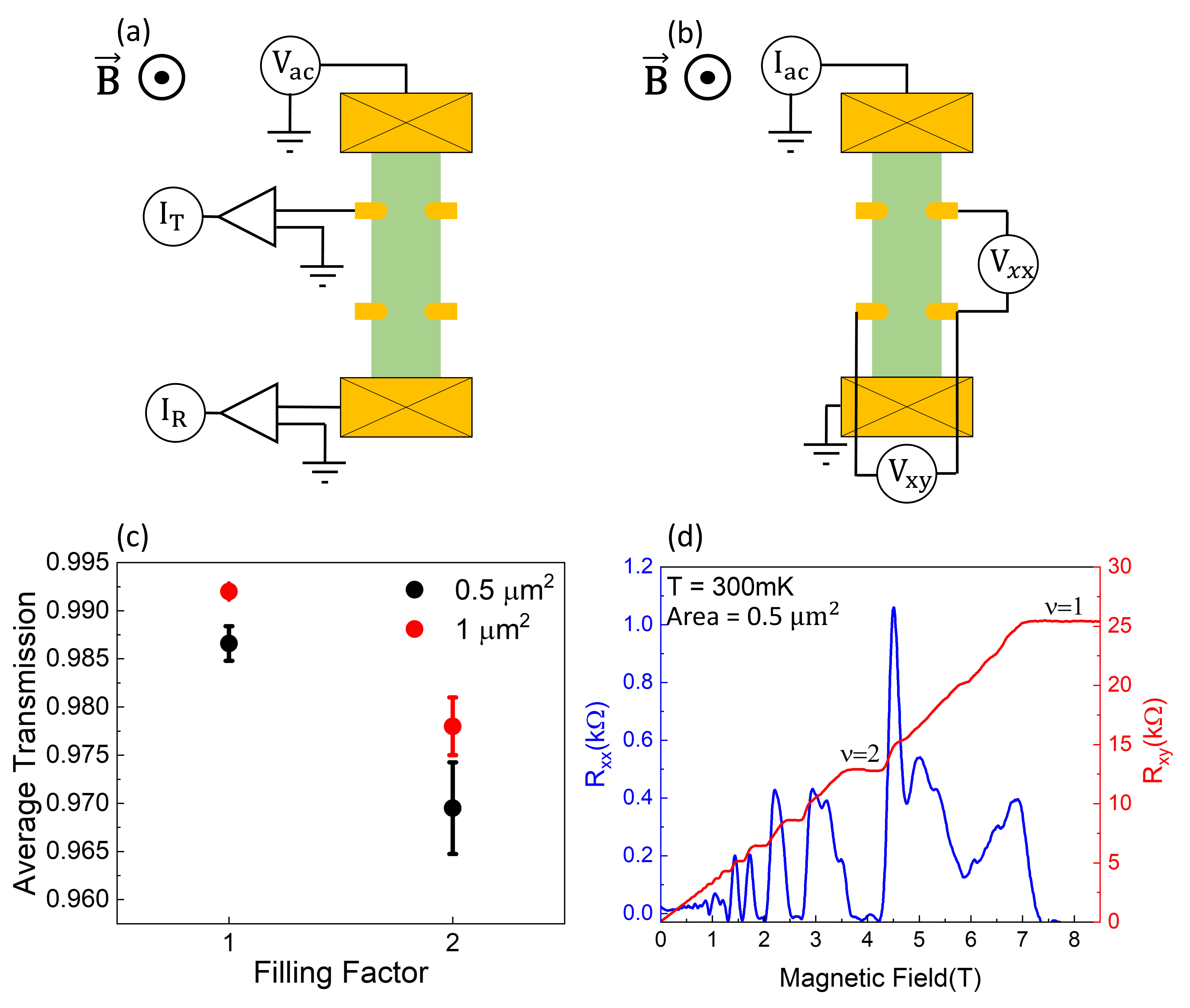}	
\caption{
 (a) Measurement configuration for assessing transmission through a micron-scale ohmic contact. 
 (b) Measurement configuration used to measure the Hall resistance and longitudinal resistance with micron-scale ohmic contacts.
 (c) Plot of the average transmission with mean error at the \(\nu\)=1,2 plateaus. All contacts used were circular and oriented along the [01\(\overline{1}\)] direction.
(d) Hall resistance and longitudinal resistance measured using micron-scale contacts as voltage probes.} 
\end{figure}
\begin{figure}
\centering 
\includegraphics[width=0.5\textwidth, angle=0]{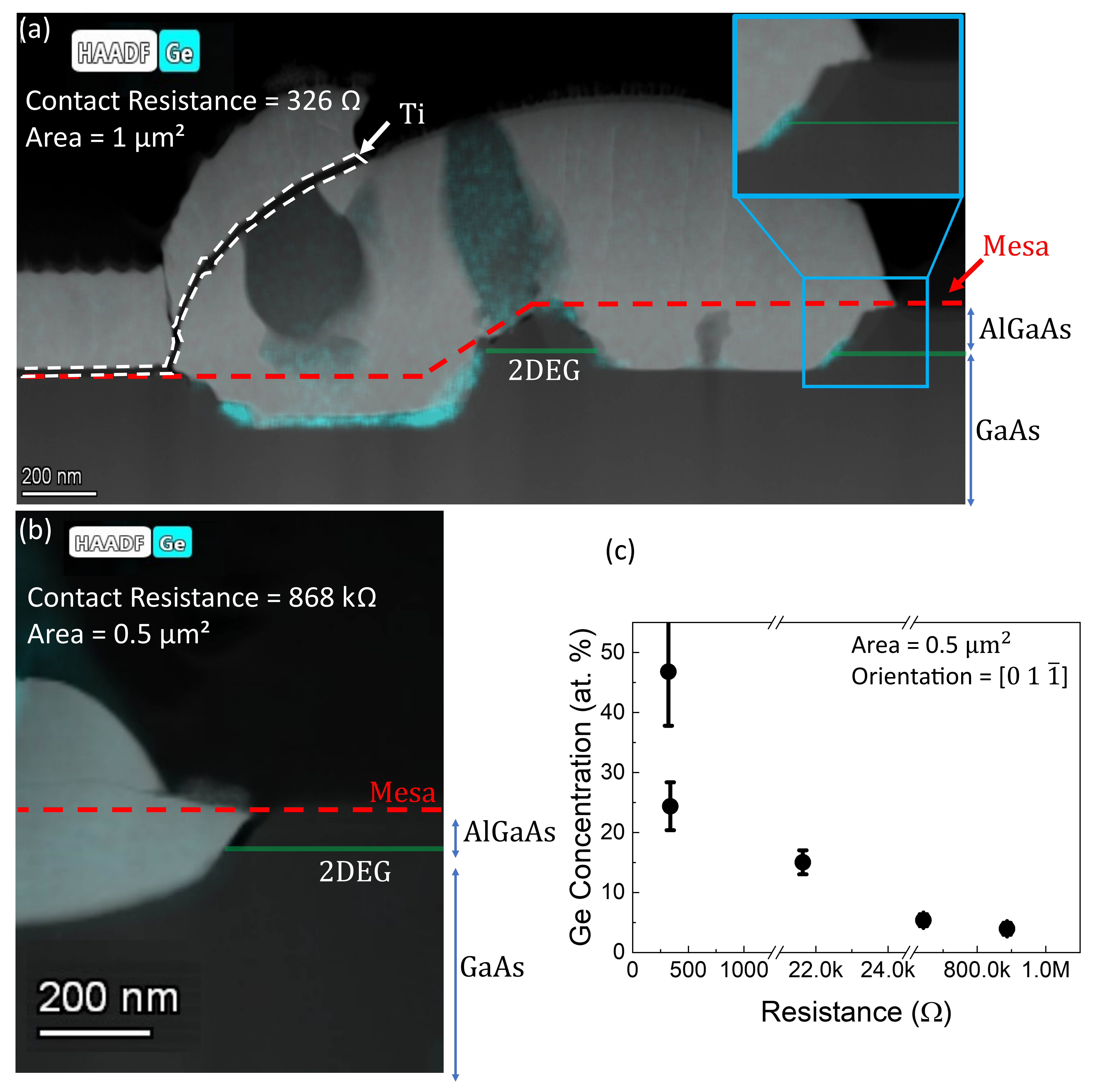}	
\caption{(a) STEM image of micron-scale ohmic contact with EDX showing high concentration of germanium at the interface of a low resistance contact. The image is a profile through the center of an annealed contact. The inset displays a close up of the interface region between the annealed metal and semiconductor.
(b) STEM image of micron-scale ohmic contact with EDX demonstrating low concentration of germanium near the metal-semiconductor interface of a high resistance contact.
(c) Graph of the concentration of germanium at the metal-semiconductor interface as a function of resistance. Each data point represents a single 0.5$\mu$m$^2$ contact. The resistance was measured at T=4.2K.}
\end{figure}

To better understand the microstructure generated in our annealed contacts, we implement scanning transmission electron microscopy (STEM) and energy-dispersive x-ray spectroscopy (EDX) to quantify local chemical composition. 
In Fig. 5(a,b) we display the concentration and location of germanium in two representative contacts following an anneal. In the low resistance contact shown in Fig. 5(a), the concentration of germanium is high near the interface between the annealed metal and the 2DEG. Fig. 5(b) displays the same germanium mapping for an insulating contact; no significant concentration of germanium is found near the interface. Fig. 5(c) shows the germanium concentration in this interfacial region for all contacts examined correlated with the measured resistances. When the germanium concentration reaches approximately 20\(\%\), the resistance drops by several orders of magnitude. The germanium heavily dopes the semiconductor region n-type and thins the Schottky barrier at the metal-semiconductor interface allowing electrons to readily tunnel into the 2DEG, and thus, the resistance drops \citep{KoopEtAl, Higmann, Saravanan}. Our microstructural analysis and electrical characterization indicate a strong correlation between the presence of germanium near the metal-2DEG interface and low resistance. All low resistance contacts have two important qualities. First, the metal spikes down into the heterostructure past the 2DEG during the anneal. Second, the area near the ohmic metal-2DEG interface becomes highly doped with germanium. Note that low germanium content near the interface as shown in Fig. 5(b) results in insulating contacts. 
In addition to germanium migration, our EDX analysis also demonstrates aluminum migration. When the ohmic metal spikes down into the heterostructure, it displaces aluminum from the AlGaAs layer towards the surface, as illustrated in Fig. 1(b)ii. Once in contact with air, the aluminum forms an oxide. This thin oxide forms an insulating barrier to any subsequent metal deposited on the contact \citep{Kamada, NissimThesis}. We use a dilute tetramethylammonium hydroxide (TMAH) solution to remove the AlOx before depositing a layer of Ti/Au on top of the contact, as shown in Fig. 1(b)iii. 

In conclusion, we have explored the parameters that control the formation of high yield, low resistance, submicron ohmic contacts to buried 2DEGs in AlGaAs/GaAs heterostructures. The local distribution of germanium at the metal-2DEG interface is demonstrated to be critically important. Furthermore, our work has shown that contact geometry and crystallographic orientation have a significant impact on the spatial distribution of germanium. Implementation of rigorous cleaning protocols ensures high yield contacts with area 0.5$\mu$m$^2$ and resistance less than 400$\Omega$.

\appendix
\section{\large{Supplementary Material}}

Devices are fabricated on a GaAs/AlGaAs heterostructure. We cleave a 5mm$\times$5mm chip from a 2-inch wafer. This chip is rinsed under agitation in toluene, acetone, and IPA. First, we define metal alignment marks using electron beam lithography. Next, we define mesas with electron beam lithography and wet etching. The semiconductor is etched for 90s with a solution of 50:5:1 H$_2$O:H$_3$PO$_4$:H$_2$O$_2$. Electron beam resist and other particulates are removed with dioxolane and IPA. In cases where residue is difficult to remove, the sample is cleaned in TMAH solution for 30s. Once mesas are defined, we use electron beam lithography to define micron-scale ohmic contacts. After the pattern has been developed, we clean the surface with an oxygen plasma etch for 15s. Following surface cleaning, metal is deposited onto the surface. During deposition extra steps are taken to preserve the quality of the metal. Approximately 50nm is excised from the crucible before any metal is deposited onto the semiconductor surface. We deposit onto the semiconductor 5nm Ni/ 80nm Ge/ 160nm Au/ 36nm Ni. Metal deposition is followed by liftoff and sequential soaks in dioxolane and IPA. The surface is also dipped in TMAH solution for 10s before being rinsed with DI water and blow dried. We verify the semiconductor surface is completely free of organic residue with atomic force microscopy (AFM) before moving onto the next step. The chip is annealed in forming gas for 2min at 435\(^\circ\)C. The Ti/Au fanout and bond pads are defined with electron-beam lithography. Immediately before metal deposition, the surface is cleaned with oxygen plasma for 60s and etched with TMAH for 50s. The chip is rinsed with DI water and blow dried before an immediate transfer to vacuum. We deposit 20nm Ti /200nm Au. After metal liftoff, the surface is cleaned with acetone, NMP, IPA, and dioxolane.

\acknowledgements
\section{\large{Acknowledgements}} This work was supported by the U.S. Department of Energy, Office of Science, National Quantum Information Science Research Centers, Quantum Science Center.
\newline
\newline
Author to whom correspondence should be addressed: mmanfra@purdue.edu

\bibliography{References}

\end{document}